\documentclass[11pt]{article}

\usepackage[a4paper,margin=1in]{geometry}
\usepackage{graphicx}
\usepackage{booktabs}
\usepackage{array}
\usepackage{amsmath}
\usepackage{amssymb}
\usepackage[numbers,sort&compress]{natbib}
\usepackage[colorlinks=true,allcolors=blue]{hyperref}
\usepackage{placeins}


\newcommand{\dlogL}{\Delta \log {\cal L}}

\newcommand{\omH}{\Omega_{\rm H}}
\newcommand{\kap}{\kappa}
\newcommand{\geff}{\gamma_{\rm eff}}
\newcommand{\gimp}{\gamma_{\rm imp}}
\newcommand{\Msun}{M_\odot}

\title{\bf Horizon-redshift transfer in black-hole direct-wave damping}

\author{
Wen-Biao Han$^{1,2,3,*}$ and Ye Jiang$^{1}$ \\
\vspace{0.3em}
\small $^{1}$Shanghai Astronomical Observatory, Chinese Academy of Sciences, Shanghai, China\\
\small $^{2}$School of Fundamental Physics and Mathematical Sciences, Hangzhou Institute for Advanced Study, University of Chinese Academy of Sciences, Hangzhou 310024, China\\
\small $^{3}$School of Astronomy and Space Science, University of Chinese Academy of Sciences, Beijing, China\\
\small $^{*}$Correspondence: wbhan@shao.ac.cn
}

\date{\today}

\begin{document}

\begin{center}
{\LARGE\bfseries Horizon-redshift transfer in black-hole\\
direct-wave damping\par}
\vspace{1.2em}
Wen-Biao Han$^{1,2,3,*}$ and Ye Jiang$^{1}$\par
\vspace{0.4em}
{\small $^{1}$Shanghai Astronomical Observatory, Chinese Academy of Sciences, Shanghai, China\par}
{\small $^{2}$School of Fundamental Physics and Mathematical Sciences, Hangzhou Institute for Advanced Study, University of Chinese Academy of Sciences, Hangzhou 310024, China\par}
{\small $^{3}$School of Astronomy and Space Science, University of Chinese Academy of Sciences, Beijing, China\par}
{\small $^{*}$Correspondence: wbhan@shao.ac.cn\par}
\vspace{1.2em}
\today
\end{center}

\begin{abstract}
Direct waves from black-hole mergers may probe horizon dynamics, but their observed envelopes need not decay at the Kerr surface-gravity rate.  We compute the complex-frequency spin-$-2$, $\ell=m=2$ Teukolsky response and combine it with a finite-duration near-horizon source whose outgoing amplitude is suppressed by gravitational redshift.  The screened Kerr response to this finite-duration source produces an observable envelope damping $\geff<\kap$.  For GW250114, this corresponds to $\geff\simeq0.4~{\rm ms}^{-1}$, consistent with a joint H1--L1 analysis of QNM-subtracted residuals.  As a consistency check, the GW231226 remnant parameters give $\geff\simeq0.31~{\rm ms}^{-1}$, compatible with the event's residual profile.  These results identify direct-wave envelope damping as an observable of horizon-redshift transfer rather than a direct measurement of surface gravity.
\end{abstract}

\section*{Introduction}

Gravitational-wave ringdown spectroscopy normally infers black-hole horizons indirectly.  The standard observables are Kerr quasinormal modes, whose complex frequencies test the remnant mass, spin and no-hair structure, but whose excitation and damping are largely governed by the light-ring potential barrier rather than by the event horizon itself \citep{vishveshwara1970,press1971,chandrasekhar1975,teukolsky1973,leaver1985,kokkotas1999,nollert1999,berti2009,dreyer2004,berti2006,yang2012qnmgeometric,cardoso2016ringdownhorizon,berti2026spectroscopy}.  Direct waves have been proposed as a different window on the merger: a prompt, non-quasinormal component sourced by the plunge, screened by the remnant potential, and partly captured in Green-function and Backwards One Body (BOB) descriptions \citep{bardeen1972,teukolsky1973,thorne1986,mino2008plunging,zimmerman2011birth,mcwilliams2019analytical,oshita2025probing,ma2026prompt,kankani2026bobdirect,lu2026gw250114}.  Here and below, by the envelope we mean the slowly varying amplitude of the direct-wave strain after factoring out a chosen carrier phase.

Recently, an analysis of GW250114 reported a post-merger direct-wave component after subtracting ordinary quasinormal-mode ringdown \citep{lu2026gw250114}.  The residual favours a direct-wave interpretation with a phase evolution compatible with the remnant horizon frequency.  However, numerical-relativity and BOB studies argue that direct-wave carrier frequencies are not generically correlated with the horizon frequency and that the GW250114 agreement may occur near an accidental spin crossing.  They also find that filtered-strain damping can evolve rather than track the surface gravity across comparable-mass simulations \citep{kankani2026bobdirect,kankani2026notmeaningful}.  This makes it essential to separate the carrier-frequency question from the envelope-damping question.  Although the GW250114 analysis in Ref.~\citep{lu2026gw250114} interpreted the direct-wave envelope as a probe of the Kerr surface gravity, its Fig.~4 shows support for finite-window damping around $0.4$--$0.5~{\rm ms}^{-1}$, below the bare Kerr value near $0.6~{\rm ms}^{-1}$.  This is not a small bookkeeping issue.  It asks whether the observed envelope should be compared with $\kap$ itself or with the screened, source-convolved signal that reaches the detector.

We argue that the reduced envelope damping is a physical consequence of the same near-horizon structure that makes direct waves interesting.  We treat $m\omH$ as a horizon-frequency reference rather than identifying it with the observed carrier.  The complex Kerr screening zero remains at $m\omH-i\kap$, while the real carrier $\omega_c$ can be supplied by the merger dynamics and varied independently.  The strain at infinity is the screened Kerr response to a finite-duration source whose outgoing amplitude is attenuated by near-horizon redshift.  Because the source amplitude and screened response decay at comparable rates, their combination produces a non-exponential envelope whose finite-time logarithmic slope can lie below $\kap$.  The observed envelope is therefore characterized by an effective damping rate $\geff$, not by $\kap$ directly.

\begin{figure}[t]
  \centering
  \includegraphics[width=0.98\linewidth]{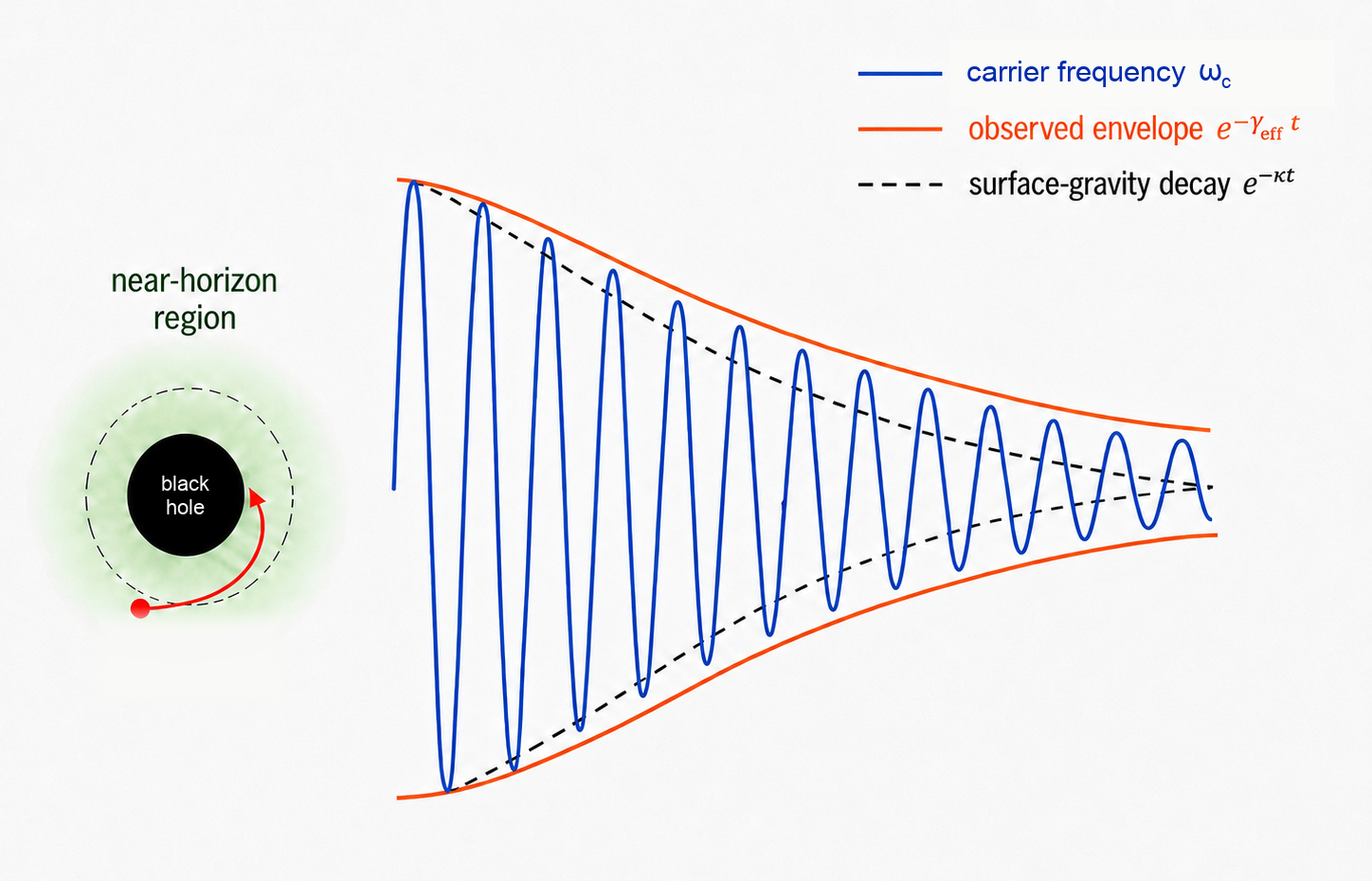}
  \caption{\textbf{Physical picture of direct-wave envelope damping.}  The direct-wave strain can be viewed as a rapidly oscillating carrier multiplied by a slowly varying envelope.  The real carrier frequency $\omega_c$ is supplied by the merger dynamics.  The envelope measured at infinity decays with the observable rate $\geff$.  The dashed guide shows surface-gravity damping, $e^{-\kap t}$, which would fall faster than the observed envelope.}
  \label{fig:physical-picture}
\end{figure}

We test this interpretation with a Teukolsky calculation and residual consistency checks in two events.  We use the homogeneous complex-frequency Teukolsky solver of \citet{jiang2026teukolsky} to compute a screened Teukolsky response kernel, convolve it with a near-horizon plunge source, and compare the resulting waveform with the public GW250114 residual data \citep{gwosc,zenodo2026gw250114}.  The complex-frequency calculation is not another damped-sinusoid fit: it separates the complex Kerr screening zero, the real carrier and the measured source-convolved envelope.  We compute the envelope damping across remnant spin and carrier frequency, compare the resulting model locus with GW250114, and use GW231226 for a second residual-level consistency check \citep{lvk2025gwtc4update}.  The physical picture is summarized in Fig.~\ref{fig:physical-picture}: the real carrier and envelope damping are distinct parts of the signal.

\section*{Direct-wave damping mechanism}

%The starting point is empirical.  
GW250114 shows a direct-wave carrier compatible with the remnant horizon frequency, but its post-merger envelope decays more slowly than expected from the bare surface-gravity rate \citep{lu2026gw250114,kankani2026notmeaningful}.  For the GW250114 remnant scale used in our fixed-parameter analysis, $M_{\rm det}\simeq68\,\Msun$ and $\chi\simeq0.68$, the Kerr horizon pole and the corresponding real horizon-frequency reference in the dominant $m=2$ channel are
\begin{equation}
  \omega_{\rm H}^{\rm pole}=m\omH-i\kap,\qquad
  f_{\rm H}\simeq190~{\rm Hz},\qquad
  \kap\simeq0.63~{\rm ms}^{-1}.
\end{equation}
If the observed direct-wave envelope damping were set simply by the imaginary part of this pole, the reference damping rate would be $\kap$.  Instead, the GW250114 residual support extends around $0.4$--$0.5~{\rm ms}^{-1}$ \citep{lu2026gw250114}, and our joint H1--L1 residual analysis of the public QNM-filtered residuals also prefers a lower envelope damping.  This residual-level comparison uses the rational-filter approach developed for black-hole ringdown mode cleaning \citep{ma2022qnmfilters,ma2023modecleaning,lu2025statistical}.  Thus, for this event, a carrier compatible with the horizon-frequency reference coexists with an envelope damping below the bare surface-gravity rate; the calculation below does not assume that the carrier agreement is universal.

\subsection*{Numerical calculation of direct-wave envelope damping}

We first calculate the direct-wave envelope damping.  The complex-frequency calculation gives a screened, Gaussian-windowed Teukolsky response kernel before source convolution.  We denote this kernel by $I(t)$ and label it ``Screened kernel'' in Table~\ref{tab:evidence}.  Before the source is introduced, the screened-kernel damping remains close to $\kap$.  Applying the same finite-window log-envelope projection to this kernel gives $\gimp\simeq0.59~{\rm ms}^{-1}$ for the GW250114 remnant scale.  We then convolve $I(t)$ with a finite-duration near-horizon plunge source, denoted $S(t)$, whose outgoing amplitude is suppressed by gravitational redshift.  Applying the same projection to the resulting direct wave gives $\geff\simeq0.39$--$0.41~{\rm ms}^{-1}$, well below $\kap$.

Table~\ref{tab:evidence} lists the calculated envelope-damping rates for the screened kernel before source convolution, the instantaneous-source limit and two finite-duration plunge-source descriptions.  In the instantaneous-source limit, an intentionally nonphysical case in which the source duration is much shorter than $\kap^{-1}$, the source-convolved waveform recovers $\gimp\simeq0.59~{\rm ms}^{-1}$.  For a finite-duration phenomenological plunge source, source convolution reduces the damping to $\geff\simeq0.41~{\rm ms}^{-1}$.  A near-horizon test-particle source gives $\geff\simeq0.40~{\rm ms}^{-1}$, providing an independent finite-duration realization.  Thus both source descriptions give envelope damping near $0.4~{\rm ms}^{-1}$.

\begin{table}[!htbp]
\caption{Calculated envelope-damping rates for the screened Teukolsky response kernel before source convolution, the intentionally nonphysical instantaneous-source limit and two finite-duration plunge-source descriptions.}
\label{tab:evidence}
\centering
\small
\begin{tabular}{@{}>{\raggedright\arraybackslash}p{0.24\linewidth}>{\raggedright\arraybackslash}p{0.20\linewidth}>{\raggedright\arraybackslash}p{0.47\linewidth}@{}}
\toprule
Case & Result & Interpretation\\
\midrule
Screened kernel & $\gimp\simeq0.59~{\rm ms}^{-1}$ & Screened Teukolsky response kernel before the source is introduced; its damping remains close to $\kap$\\
Instantaneous-source limit & $\simeq0.59~{\rm ms}^{-1}$ & Intentionally nonphysical short-duration source; recovers the screened-kernel damping\\
Plunge source & $\simeq0.41~{\rm ms}^{-1}$ & Finite-duration phenomenological source convolution reduces the envelope damping below the screened-kernel value\\
Test-particle source & $\simeq0.40~{\rm ms}^{-1}$ & A near-horizon test-particle model provides an independent finite-duration realization with nearly the same damping\\
\bottomrule
\end{tabular}
\end{table}

\begin{figure}[!htbp]
  \centering
  \includegraphics[width=0.82\linewidth]{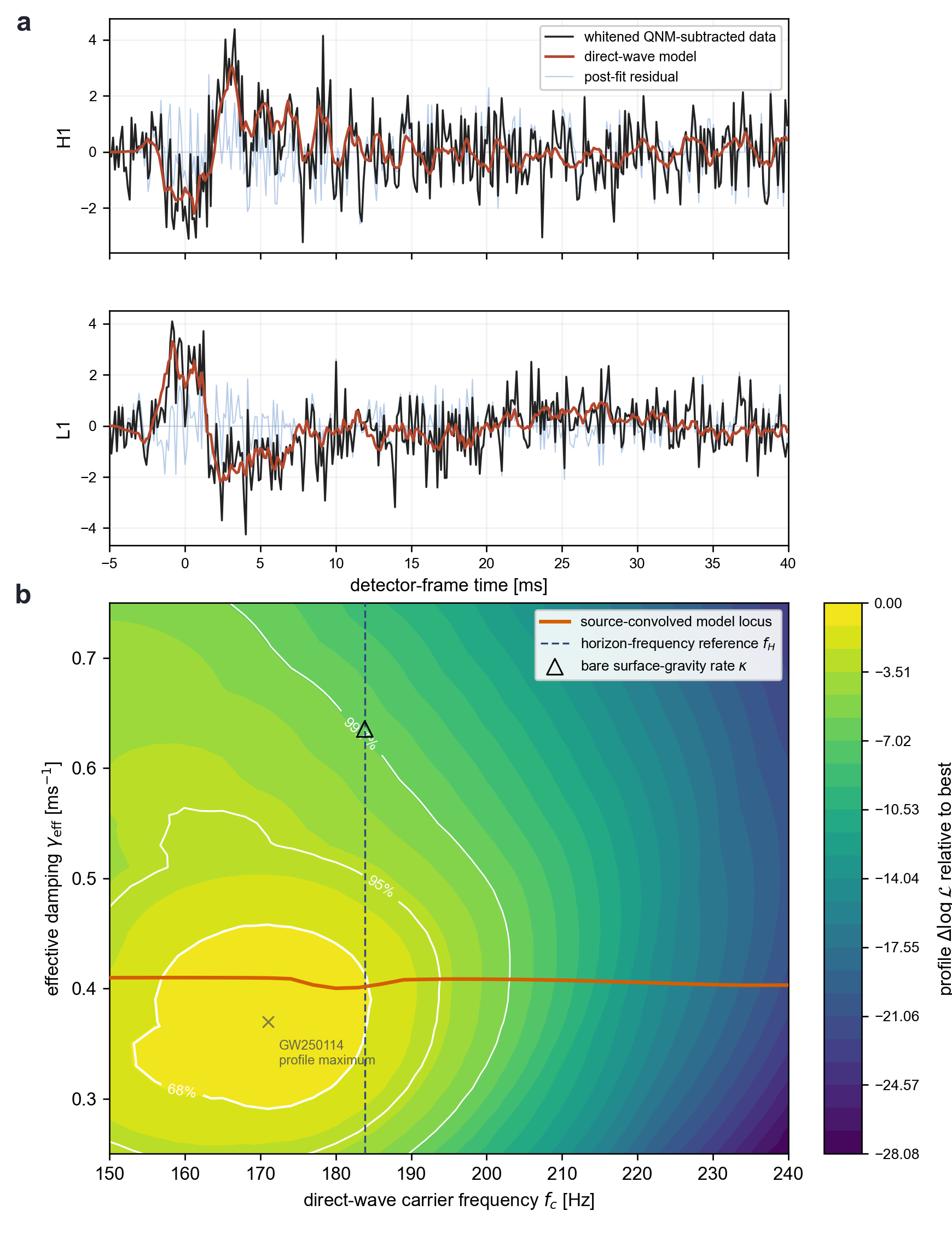}
  \caption{\textbf{GW250114 residual comparison and damping profile.}  \textbf{a,} Whitened QNM-subtracted GW250114 residuals (black) in H1 (top) and L1 (bottom), together with the source-convolved direct-wave model (red) and the post-fit residual (blue).  \textbf{b,} Joint H1--L1 profile $\dlogL$, relative to its maximum, over carrier frequency $f_c$ and effective damping $\geff$.  The orange curve is the carrier-dependent source-convolved model locus; it intersects the nominal 68\% profile region.  The dashed vertical line marks the horizon-frequency reference $f_{\rm H}$, the open triangle marks the corresponding bare-$\kap$ rate, and the cross marks the profile maximum.  White contours indicate the nominal 68\%, 95\% and 99.7\% two-parameter profile regions.}
  \label{fig:gw250114-data}
\end{figure}

Figure~\ref{fig:gw250114-data}a compares the source-convolved direct-wave signal projected into H1 and L1 with the corresponding QNM-subtracted residuals.  In the profile-likelihood map over carrier frequency and effective damping, the carrier-dependent source-convolved locus intersects the nominal 68\% profile region.  Its best point is at $f_c=171~{\rm Hz}$ and $\geff=0.410~{\rm ms}^{-1}$, only $0.18$ below the maximum in profile log likelihood.  The bare-$\kap$ reference at $f_{\rm H}$ lies outside the nominal 95\% contour.  The residual comparison therefore favours envelope damping below $\kap$ without requiring $f_c=f_{\rm H}$.

\subsection*{Physical mechanism of the reduced damping}

\begin{figure}[!htbp]
  \centering
  \includegraphics[width=0.98\linewidth]{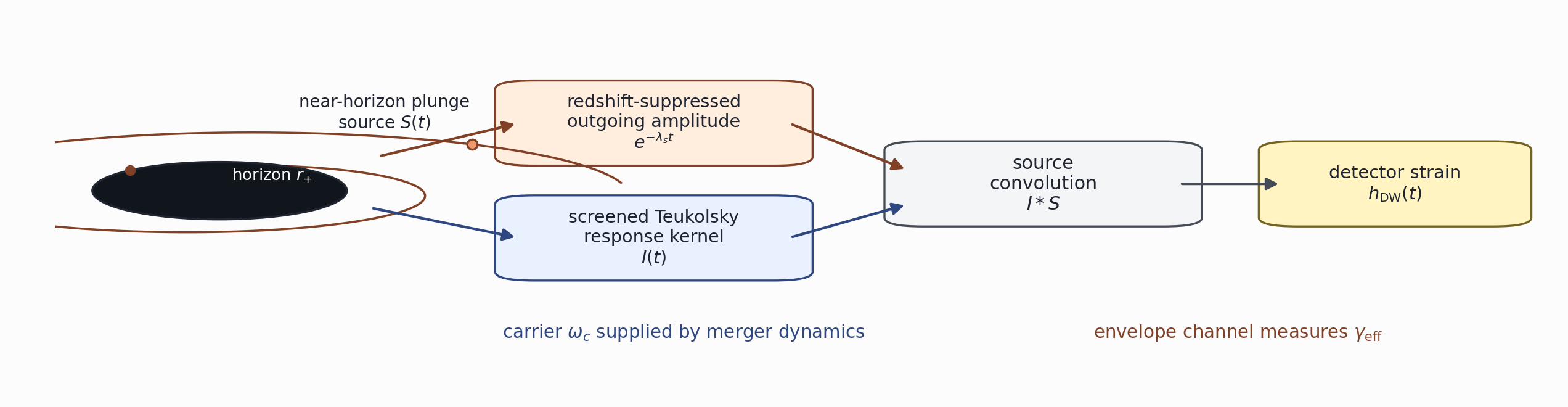}
  \caption{\textbf{Source--response mechanism for direct-wave damping.}  The screened Teukolsky response kernel $I(t)$ maps a finite-duration near-horizon plunge source $S(t)$, whose outgoing amplitude is suppressed by gravitational redshift, to the direct-wave strain $h_{\rm DW}(t)$.  The parameter $\lambda_s$ controls the exponential time dependence of $S(t)$, while the strain has real carrier $\omega_c$ and finite-window envelope damping $\geff$.}
  \label{fig:mechanism}
\end{figure}

The calculations above show that the lower damping appears when the finite-duration near-horizon source is combined with the screened kernel.  Gravitational redshift suppresses the outgoing source amplitude as the plunge approaches the horizon, while the screened Kerr response determines how this radiation reaches infinity.  Because the source and response decay at comparable rates, the resulting envelope is not a single exponential.  The finite post-peak window does not cause this broadening; it specifies the effective slope used to summarize it.  Figure~\ref{fig:mechanism} shows how the source and screened response combine in the observed direct wave.  Real-frequency template scans and time-domain damped-sinusoid fits can test consistency with the data, but do not identify these physical ingredients separately.

This physical picture can be written as a two-rate convolution.  The screening zero remains at the complex horizon frequency $\omega_{\rm H}^{\rm pole}=m\omH-i\kap$, whereas the reduced time-domain convolution uses the real carrier $\omega_c=m\omH+\delta_c\kap$.  The screening factor determines how the near-horizon Kerr response reaches infinity; it does not require the observed envelope to decay at the imaginary part $\kap$.  The parameter $\gimp$ is the damping obtained by applying the same finite-window envelope fit to the screened kernel $I(t)$ alone.  The parameter $\lambda_s$ controls the exponential time dependence of the source function $S(t)$ in observer time.  With these definitions,
\begin{equation}
  I(t)=e^{-\gimp t}e^{-i\omega_c t},\qquad
  S(t)=e^{-\lambda_s t}e^{-i\omega_c t},\qquad
  \delta_c=\frac{\omega_c-m\omH}{\kap} .
  \label{eq:two-rate-source}
\end{equation}
Their direct-wave strain is therefore
\begin{equation}
  h_{\rm DW}(t)=\int_0^t I(t-t')S(t')\,dt'
  \propto
  e^{-i\omega_c t}
  \frac{e^{-\gimp t}-e^{-\lambda_s t}}{\lambda_s-\gimp}.
\end{equation}
These expressions make the physical distinction between carrier and envelope explicit.  The common carrier controls the phase, whereas the envelope is set by the difference of two exponentials with rates of order $\kap$.  The source prescription is motivated by near-horizon redshift \citep{ma2026prompt}.  Let $x=(r-r_+)/r_+$ measure the radial distance from the outer horizon.  In Boyer--Lindquist time, a finite-energy plunge has $dx/dt\simeq-2\kap x$, so that $x\propto e^{-2\kap t}$, while a lapse-like redshift factor scales as $\sqrt{x}\propto e^{-\kap t}$.  We therefore set $\lambda_s\simeq\kap$ in $S(t)$.  The screened Kerr response then maps this near-horizon source evolution into the observed envelope damping; we refer to this mapping as horizon-redshift transfer.

The two-rate result becomes an observable damping only after applying the same finite-window log-envelope projection used in the residual analyses.  With $u=\kap t$, $\alpha_{\rm imp}=\gimp/\kap$ and $\alpha_s=\lambda_s/\kap$, the common carrier drops out of the envelope and the projected damping is
\begin{equation}
  \frac{\geff}{\kap}=
  -\frac{{\rm Cov}_{[u_1,u_2]}\left(u,\log |e^{-\alpha_{\rm imp}u}-e^{-\alpha_su}|\right)}
  {{\rm Var}_{[u_1,u_2]}(u)} .
  \label{eq:damping-relation}
\end{equation}
Equation~(\ref{eq:damping-relation}) shows how the redshift-suppressed finite-duration source modifies the screened response.  Here $\alpha_s=\lambda_s/\kap$ and $\alpha_{\rm imp}=\gimp/\kap$, while $[u_1,u_2]$ denotes the finite fitting window.  We use $\alpha_s=1$, $\alpha_{\rm imp}\simeq0.94$ and $[u_1,u_2]=[1.26,5.04]$; $\alpha_s$ follows from the redshift scaling above, while $\alpha_{\rm imp}$ and the fitting interval are determined in Methods.  The convolution produces a difference of two exponentials with rates of order $\kap$ rather than a single $e^{-\kap t}$ decay.  This broader envelope exists before the finite-window projection, which only assigns it the effective damping $\geff$.  In the equal-rate limit, the envelope becomes $t e^{-\kap t}$, with instantaneous logarithmic damping $\kap-1/t<\kap$ at finite time.  This limit shows directly why a redshift-suppressed finite-duration source can produce an observed damping below the surface-gravity rate.

For a specified screened response, source prescription and observation window, the ratio $\geff/\kap$ quantifies how the near-horizon redshift dependence enters the measured envelope damping; it is not an intrinsic decay rate of the horizon.  With the values above, Eq.~(\ref{eq:damping-relation}) gives $\geff/\kap\simeq0.62$; the spin- and carrier-dependent results below are obtained from the full numerical kernel.

\begin{figure}[!htbp]
  \centering
  \includegraphics[width=0.98\linewidth]{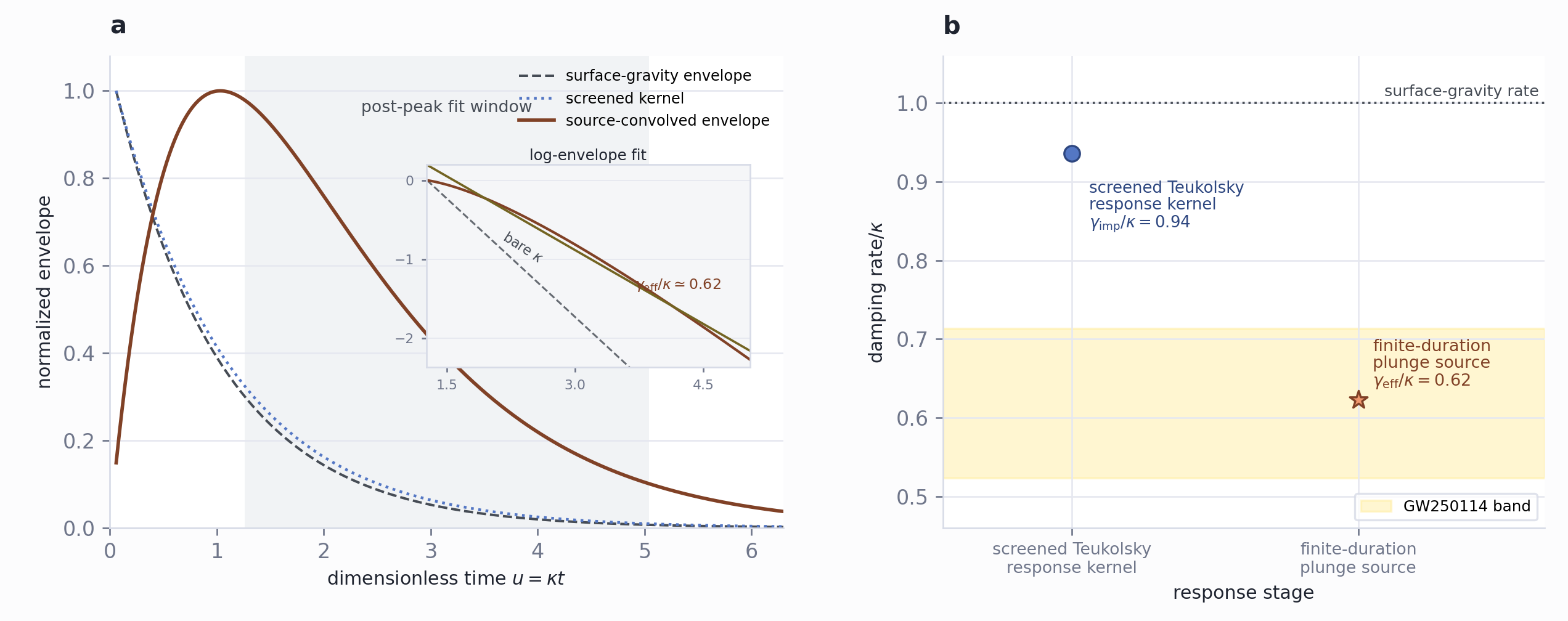}
  \caption{\textbf{Reduced direct-wave envelope damping from a finite-duration near-horizon source.}  \textbf{a,} The source-convolved envelope is broader than either a surface-gravity envelope or the screened kernel in the post-peak fit window.  The inset shows the same window in log-envelope space: the source-convolved fit has slope $\geff/\kap\simeq0.62$, shallower than the bare-$\kap$ reference line.  \textbf{b,} The blue point is the screened Teukolsky response kernel, close to the bare-$\kap$ reference rate.  The finite-duration plunge source with redshift-suppressed outgoing amplitude gives the lower observable envelope damping, $\geff/\kap\simeq0.62$.}
  \label{fig:finite-window}
\end{figure}

Figure~\ref{fig:finite-window} displays the same mechanism as a numerical envelope projection.  In the post-peak fit window, the source-convolved envelope is broader than either the bare surface-gravity envelope or the screened kernel, and its log-envelope slope gives $\geff/\kap\simeq0.62$.  If the source duration were much shorter than $\kap^{-1}$, the instantaneous-source limit would recover the screened-kernel damping near $\kap$; for the finite-duration plunge source used here, the same finite-window procedure gives a lower slope of order $0.6\kap$ in the GW250114-scaled window.

Repeating the full calculation with $\omega_c=m\omH$ over a uniform remnant-spin grid, we find $0.623\leq\geff/\kap\leq0.663$ for $0.50\leq\chi_f\leq0.95$, with the ratio remaining near $0.63$ at the spins of GW231226 and GW250114 (Fig.~\ref{fig:spin-residual}a).  Varying the real carrier independently over $-2\leq(\omega_c-m\omH)/\kap\leq1$, while holding the source prescription and dimensionless fitting window fixed, gives $0.589\leq\geff/\kap\leq0.663$, so the damping remains below $\kap$ throughout the range.  For the detector-frame remnant scales used here, the calculation with $\omega_c=m\omH$ gives $\geff=0.398~{\rm ms}^{-1}$ for GW250114 and $0.311~{\rm ms}^{-1}$ for GW231226, while the best points along the carrier-dependent event loci give $0.410$ and $0.301~{\rm ms}^{-1}$, respectively.

\FloatBarrier

\section*{Residual consistency in GW231226}

We use GW231226 for a second-event residual-level consistency test of the numerical calculation above.  It is the second-highest-SNR event in GWTC-4.0, with catalogued network signal-to-noise ratio ${\rm SNR}=33.7^{+0.1}_{-0.1}$ \citep{lvk2025gwtc4update}, and has public 4096-Hz H1 and L1 GWOSC strain together with public parameter-estimation samples.  These data allow us to propagate the remnant parameters and detector response through the same residual analysis used for GW250114.

\begin{figure}[!htbp]
  \centering
  \begin{minipage}[t]{0.40\linewidth}
    \textbf{a}\par\vspace{0.15em}
    \includegraphics[width=\linewidth,trim=0 0 275 15,clip]{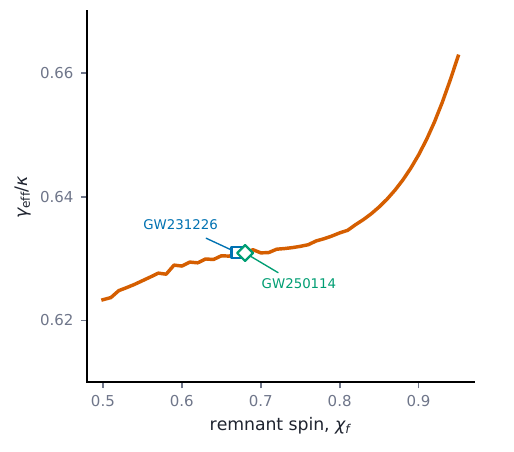}
  \end{minipage}\hfill
  \begin{minipage}[t]{0.59\linewidth}
    \textbf{b}\par\vspace{0.15em}
    \includegraphics[width=\linewidth]{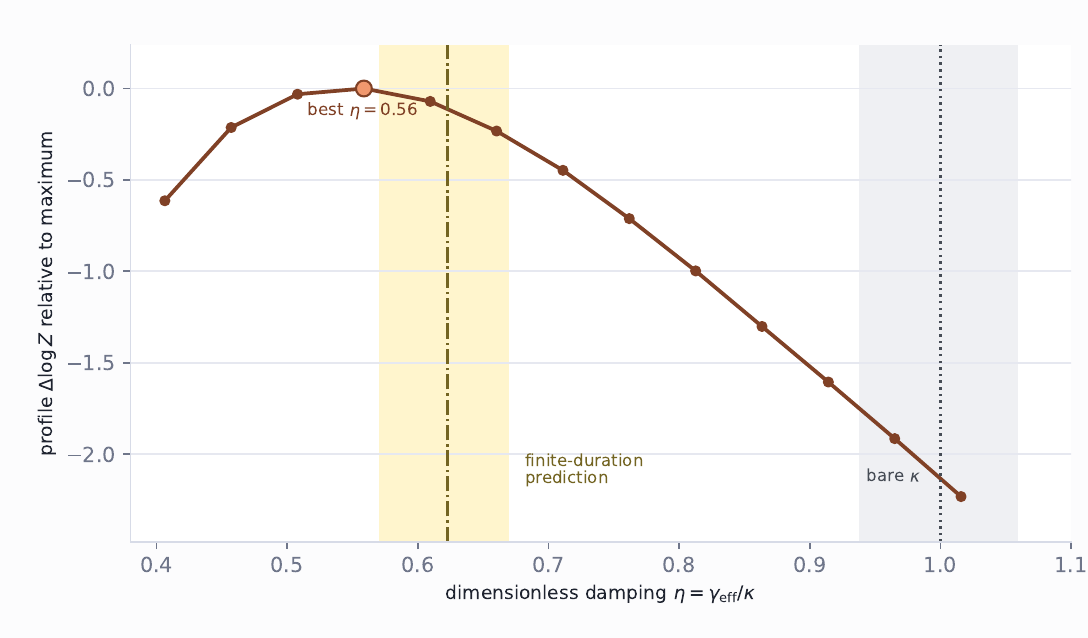}
  \end{minipage}
  \caption{\textbf{Envelope damping versus remnant spin and residual consistency in GW231226.}  \textbf{a,} Finite-window envelope damping $\geff/\kap$ as a function of remnant spin $\chi_f$, with the real carrier set to $\omega_c=m\omH$.  The open square and diamond mark the remnant spins of GW231226 and GW250114, respectively; the bare-$\kap$ reference $\geff/\kap=1$ lies outside the displayed range.  \textbf{b,} Profile log evidence $\Delta\log Z$, relative to its maximum, as a function of $\eta=\geff/\kap$ after profiling the GW231226 residual over carrier frequency.  The yellow band marks the finite-duration prediction from Eq.~(\ref{eq:damping-relation}), and the dash-dotted line marks its central value $\eta\simeq0.62$; the grey band marks the bare-$\kap$ reference range.  The profile peaks at $\eta\simeq0.56$, and the numerical value $\eta\simeq0.63$ from \textbf{a} lies within its broad maximum.}
  \label{fig:spin-residual}
\end{figure}

The GWTC-4.0 catalogue reports a source-frame remnant mass near $M_f^{\rm src}=71.4\,\Msun$ for GW231226 \citep{lvk2025gwtc4update}.  For the horizon-frequency reference and damping scales we propagate the corresponding detector-frame remnant samples, because the timescale measured at the detector is redshifted by $(1+z)$.  These samples have $M_f^{\rm det}=87.8^{+3.4}_{-3.2}\,\Msun$, $z=0.230^{+0.041}_{-0.063}$ and final spin $\chi_f=0.670^{+0.041}_{-0.036}$, where the intervals are the 5th--95th posterior percentiles.  Propagating them through the Kerr horizon formula and interpolating the numerical curve evaluated at $\omega_c=m\omH$ gives
\begin{equation}
  f_{\rm H}\simeq140^{+8}_{-7}~{\rm Hz},\qquad
  \kap\simeq0.49\pm0.03~{\rm ms}^{-1},\qquad
  \geff\simeq0.31\pm0.02~{\rm ms}^{-1},
\end{equation}
where the intervals are the 5th--95th posterior percentiles.
The numerical curve evaluated at $\omega_c=m\omH$, from which this event-level prediction is obtained, is shown in Fig.~\ref{fig:spin-residual}a.

We then profile the residual over carrier frequency and express the damping rate as $\eta=\geff/\kap$ (Fig.~\ref{fig:spin-residual}b).  The profile peaks at $\eta\simeq0.56$, while the calculation with $\omega_c=m\omH$ gives $\eta\simeq0.63$ for the GW231226 remnant.  This value lies within the broad maximum of the damping profile.  In the two-dimensional frequency--damping map, the carrier-dependent model locus also intersects the nominal 68\% region; its best point is $f_c=165.4~{\rm Hz}$ and $\geff=0.301~{\rm ms}^{-1}$, with $\eta=0.640$.  The consistency therefore does not depend on identifying the carrier with the horizon-frequency reference.

Because GW250114 and GW231226 have nearly the same remnant spin, their residual comparisons test the predicted damping scale near $\chi_f\simeq0.67$ and its conversion to different physical rates through the remnant mass; they do not independently measure the spin dependence shown in Fig.~\ref{fig:spin-residual}a.

\FloatBarrier

\section*{Discussion}

The damping measured in a direct-wave strain envelope is not generically the rate $\kap$ set by the Kerr surface gravity.  In the conditional Kerr-screened interpretation considered here, the observed envelope is shaped by the finite-duration source and screened response and should not be identified with the imaginary part of the complex horizon-frequency reference.  The real carrier may be supplied by the merger dynamics, while the complex screening zero retains its dependence on $m\omH$ and $\kap$.  In the adopted source model, near-horizon redshift suppresses the outgoing plunge radiation, and the screened Kerr response maps that radiation to infinity.  Their comparable decay rates produce a non-exponential envelope, which is summarized by $\geff$ over a specified measurement window.  The ratio $\geff/\kap$ is therefore conditional on the source prescription and window rather than a universal horizon coefficient.

Recent numerical-relativity results sharpen this distinction.  The filtered direct-wave carrier departs systematically from $2\omH$ away from remnant spins near $0.7$, while the extracted imaginary frequency varies substantially over short time intervals \citep{kankani2026notmeaningful}.  Those time-dependent ranges are not a direct measurement of the finite-window envelope slope used here, but they show why a single fitted damping rate should not be identified with $\kap$ without specifying the source and measurement window.

The numerical calculation gives $0.589\leq\geff/\kap\leq0.663$ across the remnant-spin and real-carrier ranges considered.  For $\omega_c=m\omH$, the ratio increases gradually from $0.623$ to $0.663$.  GW250114 and GW231226 both have remnant spins near $0.67$, so their residual agreement tests the characteristic damping scale and its mass rescaling in this part of parameter space, not the calculated spin trend.  The value near $0.6\kap$ is conditional on the finite-duration source prescription and dimensionless measurement window, rather than a universal constant; it can shift with either choice, and in the impulsive limit the fitted damping approaches the screened-kernel rate near $\kap$.

The source-convolved calculation shows how the observed strain envelope can carry an imprint of the underlying Kerr surface gravity $\kap$.  Because the carrier need not equal the horizon frequency, the envelope should be interpreted through the screened Kerr response and source evolution rather than through the carrier phase alone.  With higher-SNR mergers and joint inference of the remnant and source, $\kap$ may be inferred from the measured damping without identifying $\geff$ with $\kap$.  Tests across masses, spins and source configurations could ultimately establish whether black-hole horizon properties can be inferred directly from gravitational-wave data.

\section*{Methods}

\subsection*{Complex-frequency screened response}

We solve the spin-$-2$, $\ell=m=2$ homogeneous radial Teukolsky equation using the analytic-series numerical method of \citet{jiang2026teukolsky}.  The radial solutions and asymptotic amplitudes are evaluated at complex $\omega$, so that the carrier frequency and damping rate enter the same Kerr transfer function.  The screened Teukolsky response kernel is then computed before the near-horizon plunge source is introduced.

For each complex $\omega$, we construct the ingoing and upgoing homogeneous radial solutions.  The ingoing solution $R^{\rm in}$ is purely ingoing at the outer horizon and becomes a superposition of ingoing and outgoing waves at infinity.  The upgoing solution $R^{\rm up}$ is purely outgoing at infinity and becomes a superposition of ingoing and outgoing waves at the horizon.  Following the Jiang--Han scheme, the solutions are represented by Frobenius or asymptotic series near the singular points and by ordinary-point power series between them.  The series coefficients obey recurrence relations, and matching across overlapping radial intervals yields the asymptotic amplitudes and the Wronskian.  This construction avoids direct long-range radial integration at each sampled frequency.

For each complex $\omega$, the transfer-function calculation records $C^{\rm trans}$, $B^{\rm inc}$ and $W=2i\omega B^{\rm inc}$.  The direct-wave screening factor vanishes at the first complex horizon frequency \citep{zimmerman2011birth,oshita2025probing}.  We therefore represent its leading local behaviour by defining the screened response as
\begin{equation}
  G_{\rm scr}(\omega)=(\omega-\omega_{\rm H}^{\rm pole})\frac{C^{\rm trans}}{W},
  \qquad \omega_{\rm H}^{\rm pole}=m\omH-i\kap .
\end{equation}
The linear factor $(\omega-\omega_{\rm H}^{\rm pole})$ ensures that $G_{\rm scr}$ vanishes at $\omega_{\rm H}^{\rm pole}$.  In a full source--response integrand, this zero screens a coincident horizon-mode pole; it does not imply that $C^{\rm trans}/W$ itself contains that pole.  We evaluate $G_{\rm scr}$ along the contour ${\rm Im}\,\omega=-\kap$, where the factor reduces to ${\rm Re}\,\omega-m\omH$.  The Gaussian frequency window is centred on the real carrier $\omega_c$ before transformation to the time domain.  To test the dependence on the real carrier, we shift this window and the source carrier together over $-2\leq(\omega_c-m\omH)/\kap\leq1$, while keeping the complex screening zero fixed.  The common imaginary part of the contour supplies an $e^{-\kap t}$ factor, while the screened transfer function and frequency window determine the finite-time shape of the kernel.  We denote the resulting screened, Gaussian-windowed Teukolsky response kernel by $I(t)$ and obtain $\gimp$ by applying the finite-window envelope fit to this kernel.  The near-horizon source defined in Eq.~(\ref{eq:two-rate-source}) is applied separately in the subsequent time-domain convolution.  This construction gives a screened, band-limited Kerr kernel rather than a full inhomogeneous merger waveform.

\subsection*{Finite-window damping projection}

The phenomenological plunge source and near-horizon test-particle source listed in Table~\ref{tab:evidence} are each convolved with the same screened kernel and evaluated using the same finite-window projection.  We obtain the envelope damping $\geff$ from a linear fit to the log envelope over a finite dimensionless window.  With $u=\kap t$, we fit
\begin{equation}
  \log |h(u)|\simeq c-\beta u
\end{equation}
by minimizing
\begin{equation}
  \int_{u_1}^{u_2}\left[\log |h(u)|-c+\beta u\right]^2du .
\end{equation}
The least-squares slope is the covariance projection given in Eq.~(\ref{eq:damping-relation}); for the numerical calculation, the analytic two-rate envelope inside the logarithm is replaced by the full waveform envelope $|h(u)|$.  The covariance and variance are uniform-window averages, multiplicative normalizations and the carrier phase do not affect the slope, and the reported dimensionless damping is $\geff/\kap=\beta$.

The fitting interval is inherited from the $2$--$8~{\rm ms}$ post-peak window used for GW250114.  With $\kap=0.6306~{\rm ms}^{-1}$, its dimensionless endpoints are $u_1=\kap t_1=1.261$ and $u_2=\kap t_2=5.045$.  We retain this interval in units of $u=\kap t$ when evaluating other remnant parameters; the corresponding interval in milliseconds therefore rescales with $\kap^{-1}$ rather than remaining fixed at $2$--$8~{\rm ms}$.

The value of $\alpha_{\rm imp}$ is obtained independently from the screened response.  For the GW250114 remnant with $\omega_c=m\omH$, we calculate the screened, band-limited Kerr kernel $I(t)$ and apply the same log-envelope fit over $[u_1,u_2]$.  This gives $\gimp\simeq0.59~{\rm ms}^{-1}$ and hence $\alpha_{\rm imp}=\gimp/\kap\simeq0.94$.  Its proximity to unity follows from the common $e^{-\kap t}$ factor supplied by the complex-frequency contour, while the screened transfer function and finite frequency window modify the kernel over the fitted interval.  Together with $\alpha_s=1$ from the source prescription in Eq.~(\ref{eq:two-rate-source}), this kernel value and $[u_1,u_2]=[1.26,5.04]$ give $\geff/\kap\simeq0.62$ through Eq.~(\ref{eq:damping-relation}).

\subsection*{GW250114 residual scan}

We use the public GW250114 strain data, posterior samples and QNM-filtered residual products from refs.~\citep{lu2026gw250114,gwosc,zenodo2026gw250114}.  We evaluate the detector response using the NRSur7dq4 posterior reconstruction adopted for the public residual products \citep{varma2019nrsur}.  The plunge-source parameters used in Fig.~\ref{fig:gw250114-data}a were selected by minimizing the whitened residual sum of squares over $-8\leq t/M_{\rm det}\leq20$.  The target direct-wave component was obtained by subtracting the direct-wave-subtracted residual from the QNM-filtered residual.  We scanned start frequencies of 226, 236 and 246~Hz, frequency-relaxation times of 6, 8 and 10~ms, and source-decay times of 0.8, 1.2, 1.5 and 2.0~ms.  The source-rise time was fixed at 0.2~ms and the detuning exponent at 1.5.  At each grid point, we fitted two quadrature amplitudes linearly and selected the onset from 241 samples spanning $-12$ to $8\,M_{\rm det}$.  The minimum occurred at a start frequency of 246~Hz, a frequency-relaxation time of 10~ms and a source-decay time of 2~ms.

With these parameters fixed, we project the complex mode into H1 and L1 using the maximum-posterior detector response.  We jointly fit a complex amplitude, equivalent to two real quadrature coefficients, and a discrete time offset in the whitened residual basis.  For Fig.~\ref{fig:gw250114-data}b, we vary the carrier frequency and effective damping, profile over the amplitude and time offset, and report the profile likelihood over $(f_c,\geff)$.  With frequency in Hz and $\kap$ in ${\rm ms}^{-1}$, the model locus is obtained from $\delta_c=2\pi(f_c-f_{\rm H})/(10^3\kap)$ and interpolation of the numerical carrier--spin surface at the fixed remnant sample used for this map.

\subsection*{GW231226 residual profile}

The GW231226 residual test uses the public 4096-Hz H1 and L1 GWOSC strain files and public parameter-estimation samples.  The event time and detector response are fixed by the maximum-posterior sample, whose remnant mass and spin are also used to filter the $(2,2,0)$--$(2,2,2)$ Kerr modes from each detector strain.  The full remnant posterior is propagated separately through the Kerr relations and the numerical curve evaluated at $\omega_c=m\omH$ to obtain the quoted ranges of $f_{\rm H}$, $\kap$ and $\geff$.  At each $(f_c,\geff)$, the finite-duration source-convolved waveform is projected into H1 and L1 and compared jointly with the QNM-filtered residuals.  The joint evidence $Z$ is evaluated after marginalizing over the two quadrature amplitudes and a discrete time-offset grid.  The one-dimensional curve in Fig.~\ref{fig:spin-residual}b is obtained by profiling $Z(f_c,\geff)$ over $f_c$ and plotting $\Delta\log Z$, relative to its maximum, against $\eta=\geff/\kap_{\rm med}$, where $\kap_{\rm med}$ is the posterior median.  The two-dimensional model locus is evaluated separately at the fixed remnant sample used for the map; this map-specific sample is not the posterior median.

\section*{Data and code availability}

The derived data products and analysis code will be deposited in a versioned public repository upon publication.

\bibliographystyle{unsrtnat}
\bibliography{references_v31}

\end{document}